\documentclass[twocolumn,aps,prl,float]{revtex4}

\usepackage{graphicx}% Include figure files
\usepackage{bm}% bold math

\begin{document}

\title{Spin Nematic Phase in $S$=1 Triangular Antiferromagnets}
\author{Hirokazu Tsunetsugu}
\author{Mitsuhiro Arikawa}
\affiliation{%
Yukawa Institute for Theoretical Physics, 
Kyoto University, Kyoto 606-8502, Japan}

%\date{\today}

\begin{abstract}
Spin nematic order is investigated for a $S$=1 spin model 
on triangular lattice with bilinear-biquadratic interactions.  
We particularly studied an antiferro nematic order phase 
with three-sublattice structure, 
and magnetic properties are calculated at zero temperature 
by means of bosonization.  
Two types of bosonic excitations are found.  One is a
gapless excitation with linear energy dispersion around 
$\mathbf{k} \sim \mathbf{0}$, and this leads to 
a finite spin susceptibility at $T=0$ and would  have a 
specific heat $C(T) \sim T^2$ at low temperatures.  
These behaviors can explain many of characteristic 
features of recently discovered spin liquid state in 
the triangular magnet, NiGa$_2$S$_4$.  
\end{abstract}

% PACS, the Physics and Astronomy
                           % Classification Scheme.

\maketitle

The concept of spin liquid was introduced thirty years 
ago by P.~W.~Anderson \cite{anderson}, as a quantum critical state in 
which the spin-spin correlation function does not show a 
real long-range order but a power-law behavior.  
This issue has been studied intensively since then 
both theoretically and experimentally.  Frustration and 
quantum fluctuations are considered two ingredients to realize 
a spin liquid, and a spin-1/2 Heisenberg 
antiferromagnet on triangular lattice was the first 
candidate.  Many antiferromagnetic 
materials with triangular lattice structure have been 
studied to see if they may show a spin liquid behavior, but 
most of them turned out to exhibit some long-range 
order at low temperatures.  Very few exceptions 
are ${}^3\mathrm{He}$ thin layer \cite{He} and 
organic $\kappa$-(BEDT-TTF)$_2$Cu$_2$(CN)$_3$ \cite{BEDT},   
and just recently a new spin-liquid material NiGa$_2$S$_4$ was 
discovered \cite{nigas}. While the former two are spin-1/2 
system, NiGa$_2$S$_4$ is a spin-1 system.  

In NiGa$_2$S$_4$, spins of Ni$^{2+}$ ions 
form triangular layers with undistorted regular triangle units, 
and the layers are stacked along the c-axis.  
These layers are effectively decoupled, 
since the Ni-Ni distance is more than three times longer 
between layers.  
This system showed various low-temperature 
properties that indicate a spin liquid state.  
First of all, no singularity was observed in 
specific heat down to the lowest temperature, 
$T$=0.3 K, meaning the absence of phase transitions.  
Moreover, the specific heat shows a power-law 
behavior, $C \sim T^2$, below 10K.  
Secondly, the magnetic susceptibility gradually increased 
with decreasing temperature, and approached 
a finite value.  Thirdly, neutron experiment observed 
a peak at an incommensurate wavevector 
$\mathbf{Q}$$\sim$$(\frac{\pi}{\sqrt{3}},0)$.  
However, this was not a magnetic 
Bragg peak, and spin correlation length did not 
diverge but saturated to about $\xi$$\sim$ 20\AA, 
only seven lattice units.  

Absence of magnetic long-range order and presence of 
critical behaviors are necessary conditions to 
identify a spin liquid, and these were satisfied 
in NiGa$_2$S$_4$.  
These may suggest a finite spin gap instead, but this 
contradicts a nonvanishing temperature dependence of susceptibility.  
In this paper, we will examine the possibility 
of a hidden order that reproduce similar low-temperature 
properties as critical spin liquid states.  

%-------------------------------
\begin{figure}[bt]
\begin{center}
\includegraphics[width=7cm]{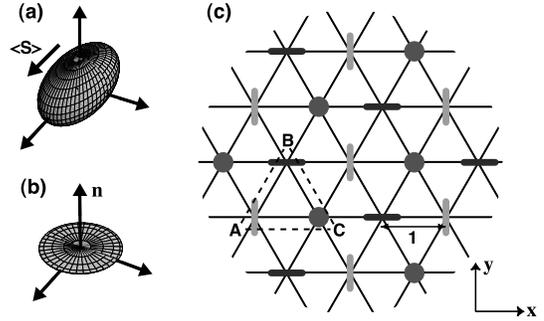}
\end{center}
\caption{Spin quadrupole moment $\mathcal{Q}_{\mu \mu'}$ 
of a single-site wavefunction 
when (a) $\langle \mathbf{S} \rangle \ne \mathbf{0}$ 
and (b) $\langle \mathbf{S} \rangle = \mathbf{0}$. 
(c) Three-sublattice nematic order.  
Dotted triangle shows a unit cell of the ordered state.}
\label{fig:order}
\end{figure}
%-------------------------------

Possible order parameters are not ordinary 
static spin dipole moments $\langle \mathbf{S} \rangle$, 
since neutron experiment did not observe magnetic Bragg peaks. 
We should note that Ni$^{2+}$ ions do not have 
orbital degrees of freedom \cite{nigas}, and 
that we can describe this system 
as a pure spin model with no spin anisotropy. 
Therefore, if any long-range order exists, its order parameter 
should be represented in terms of spin operators.
We will investigate the simplest 
candidate, spin quadrupole moments,  
$Q_{\mu \mu'} = \frac12 \langle S^{\mu} S^{\mu'} 
+ S^{\mu'} S^{\mu} \rangle 
- \frac13 S(S+1) \delta_{\mu \mu'}$, 
where $\mu$ is spin index, 
and this also corresponds to nematic order \cite{nematic,matveev}.
The nematic order parameter $Q_{\mu \nu}$ describes 
anisotropy of spin fluctuations, not 
static moment, and can be nonzero only if $S \ge 1$ 
\cite{momoi}.  
In NiGa$_2$S$_4$, local spins are $S$=1 and therefore 
we consider the nematic order parameter defined 
at each site.  
Neutron experiment found a peak of scattering at 
incommensurate wavevectors $\mathbf{Q}$, not at 
$\mathbf{q}$=$\mathbf{0}$ or Brillouin zone boundary.  
This suggests that the expected nematic order is not 
uniform but modulates in space, i.e., some antiferro 
order.  

To describe this order, the standard Heisenberg 
Hamiltonian is not sufficient, since it 
is believed to have a 120-degree magnetic 
long range order \cite{bernu}.  We use a spin-1 model 
with additional biquadratic interactions between 
nearest neighbor sites on the triangular lattice, 
\begin{equation}
  H = \sum_{\langle i,j \rangle}
  \left[ J \mathbf{S}_i \cdot \mathbf{S}_j + 
    K ( \mathbf{S}_i \cdot \mathbf{S}_j )^2 \right] .
  \label{eq:ham}
\end{equation}
This model has been studied by mean field analysis and 
numerical calculations \cite{nematic_papers}.  
The mean field analysis at $T=0$ showed two nematic 
phases.  One is the region of $K$$<$$J$$<$0 and 
a uniform nematic order appears.  The other one 
is an antiferro nematic order and it is predicted 
in the region of 0$<$$J$$<$$K$.  
The latter is our expected case, 
and we will investigate that parameter region.   
We should note that the model (\ref{eq:ham}) is a phenomenological 
Hamiltonian introduced to describe an expected 
modulated nematic order.  
While the antiferromagnetic 
bilinear terms are naturally expected from 
superexchange processes, the biquadratic terms 
are also present as higher-oder processes of virtual electron 
hopping, but usually it is expected 
that $|J|$$>$$|K|$.   
In our choice of model, it is assumed that both 
coupling constants are renormalized 
in the low-energy sector from their microscopic 
values such that 0$<$$J$$<$$K$, but this 
should be examined in future work.  

We investigate the nature of nematic 
order parameter $Q_{\mu \nu}$ in more details 
for the $S$=1 case.  To see this, it is more 
convenient not to subtract the constant term 
in the definition, and consider 
$\mathcal{Q}_{\mu \nu} (\mathbf{r})$=$\frac12 
\langle S^{\mu}  (\mathbf{r}) S^{\nu}  (\mathbf{r})$
+$S^{\nu}  (\mathbf{r}) S^{\mu}  (\mathbf{r})\rangle$ 
at site $\mathbf{r}$.  
We introduce this tensor at each site $\mathbf{r}$ 
defined for a local wavefunction, and calculate its 
average by taking account of fluctuations in space 
and time.  
In the $S$=1 case, we can show that 
for any single-site wavefunctions, 
the three eigenvalues of $\mathcal{Q}_{\mu \nu}$ are 
$\lambda_{1,2}$=$\frac12 ( 1 \mp \sqrt{1-\langle \mathbf{S} \rangle^2} )$, 
and $\lambda_3$=1, and ${\mathcal Q}_{\mu \nu}$ is represented 
as an ellipsoid as shown in Fig.~\ref{fig:order}(a).  
Therefore, when $\langle \mathbf{S} \rangle$=$\mathbf{0}$, 
spin fluctuations are like disk and have zero amplitude 
in one direction in spin space (see Fig.~\ref{fig:order}(b)).  
This is characterized 
by the director vector $\mathbf{n}$ that is 
perpendicular to the disk. The corresponding wavefunction 
$|\psi_\mathbf{n}\rangle$ is the coherence state such that 
$(\mathbf{n} \cdot \mathbf{S}) | \psi_\mathbf{n} \rangle $=0, 
namely rotated $|\mbox{$S_z$=0} \rangle$ state with quantization 
axis parallel to $\mathbf{n}$.    

Mean field solution of the model (\ref{eq:ham}) is 
easily obtained for the triangle lattice, 
and it is a three-sublattice nematic 
order shown in Fig.~\ref{fig:order}(c). 
The directors in the three sublattices are orthogonal 
to each other, and without generality, we can set 
$\mathbf{n}_A $$\parallel $$x$,  
$\mathbf{n}_B $$\parallel $$y$, and  
$\mathbf{n}_C $$\parallel $$z$, where $A$-$C$ are 
sublattice indices.  
Then the mean-field ground 
state is a direct product of local states 
$|\Psi_{\mathrm{MF}} \rangle$=
$\prod_{\mathbf{R}} | \mbox{$S_x$=0} \rangle_{A, \mathbf{R}}$
$\otimes$
$| \mbox{$S_y$=0} \rangle_{B, \mathbf{R}}$
$\otimes$
$| \mbox{$S_z$=0} \rangle_{C, \mathbf{R}}$. Here, 
$\mathbf{r}$=$(j,\mathbf{R})$, with 
sublattice index $j \in \{A,B,C \}$ and 
unit-cell coordinate $\mathbf{R}$.  

We now describe quantum fluctuations of the mean-field 
solution by introducing bosons.  
Each $S$=1 spin has three local states, 
and one of them corresponds to 
a mean-field solution. Following Matveev \cite{matveev}, 
we consider 
it as a local vacuum and the other two excited states 
as one-boson states.  To be specific, let us consider 
a site in the C-sublattice.  
$|\mbox{$S_z$=0}\rangle_{C,\mathbf{R}}$ 
is local vacuum $|\mbox{vac}\rangle_{C,\mathbf{R}}$, and 
the other states are represented as 
$|\mbox{$S_z$=$\pm$1}\rangle_{C,\mathbf{R}}$=
$2^{-1/2}$ 
$( \alpha_{C,\mathbf{R}}^\dagger $$\pm$$ i \beta_{C,\mathbf{R}}^\dagger )$ 
$|\mbox{vac}\rangle_{C,\mathbf{R}}$,  
by boson creation operators $\alpha^\dagger$ and $\beta^\dagger$.   
Then spin operators are represented as 
$S_{C \mathbf{R}}^x$=$\alpha_{C \mathbf{R}}^\dagger$+$\alpha_{C \mathbf{R}}$, 
$S_{C \mathbf{R}}^y$=$\beta_{C \mathbf{R}}^\dagger$+$\beta_{C \mathbf{R}}$, 
$S_{C \mathbf{R}}^z$=$-i (\alpha_{C \mathbf{R}}^\dagger \beta_{C \mathbf{R}}$
$-$$\beta_{C \mathbf{R}}^\dagger \alpha_{C \mathbf{R}} )$. 
In a similar way, two types of bosons are also introduced 
for the A- and B-sublattices.  
It is noted that these bosons 
are subject to constraint, 
$\alpha_{j \mathbf{R}}^\dagger \alpha_{j \mathbf{R}}$
+$\beta_{j \mathbf{R}}^\dagger \beta_{j \mathbf{R}}$$\le$1, 
or equivalently $\alpha_{j \mathbf{R}}^2$=$\beta_{j \mathbf{R}}^2$
=$\alpha_{j \mathbf{R}} \beta_{j \mathbf{R}}$=0. 

%-------------------------------
\begin{figure}[t]
\begin{center}
\includegraphics[width=6cm]{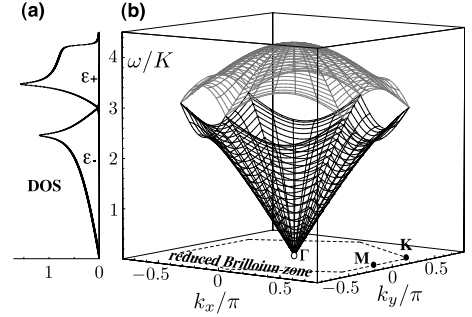}
\end{center}
\caption{Two types of 
excitations for $J/K$=0.5.(a) Density of states with logarithmically 
divergent van Hove singularity. (b) Energy dispersion.}
\label{fig:eng}
\end{figure}
%-------------------------------

We replace spin operators in the model (\ref{eq:ham}) 
by bosons and then neglect their interactions.  
This corresponds to Gaussian approximation 
of spin fluctuations. 
After taking the Fourier transformation, 
the obtained boson hamiltonian reads, 
$ H_{b}$=$3K\Omega$+ 
$ H( \beta_{A} , \alpha_{B} )$+
$ H( \beta_{B} , \alpha_{C} )$+ 
$ H( \beta_{C} , \alpha_{A} )$, 
and 
\begin{eqnarray}
  &&H( \beta_{j} , \alpha_{j'} )  =
  3K \sum_{\mathbf{k}} \left(  
  \beta_{j \mathbf{k}}^\dagger   \beta_{j \mathbf{k}} + 
  \alpha_{j' \mathbf{k}}^\dagger \alpha_{j' \mathbf{k}} \right)
\nonumber\\ 
  && + 3 \sum_{\mathbf{k}} \left\{
  \gamma_{\mathbf{k}} \left[ 
  (\mbox{$J$$-$$K$}) 
  \beta_{j \mathbf{k}}^\dagger   \alpha_{j' \mathbf{k}} 
  + J 
  \beta_{j \mathbf{k}}^\dagger   \alpha_{j' \mathbf{k}}^\dagger \right]
  + \mbox{h.c.} \right\} , 
  \label{eq:b_ham}
\end{eqnarray}
where $\Omega$ is the number of unit cells and the 
sum is taken over the reduced Brillouin zone 
of the three-sublattice order, and 
$\gamma_{\mathbf{k}}$=$\frac13 e^{-ik_x}$+
$\frac23 e^{ik_x /2} \cos (\sqrt{3}k_y /2) $.  
Here we set the lattice constant being unity.
Each type of bosons interact 
with only another one type of bosons, and therefore 
we can easily diagonalize the boson hamiltonian by 
Bogoliubov transformation, 
$H_b = \sum_{\mathbf{k}, j\in \{A,B,C\} } \sum_{m=\pm}
 \varepsilon_{m ,\mathbf{k}}
  b_{m \mathbf{k}}^\dagger b_{m \mathbf{k}} + E_0 $ 
with the eigenenergy 
\begin{equation}
  \varepsilon_{\pm , \mathbf{k}} = 3K 
  \sqrt{ ( 1 \pm  |\gamma_\mathbf{k}| ) 
   ( 1 \pm  \kappa |\gamma_\mathbf{k}| ) }  , 
  \quad \kappa = 1 - 2J/K . 
  \label{eq:eng}
\end{equation} 

The dispersions and density of states 
are plotted in Fig.~\ref{fig:eng} for $J/K$=0.5.
The $\varepsilon_-$ branch is gapless excitation with 
asymptotically linear dispersion, 
$\varepsilon_{-,\mathbf{k}} \sim 3 \sqrt{JK/2} \, |\mathbf{k}|$ 
around $\mathbf{k}$=$\mathbf{0}$.  
The $\varepsilon_+$ branch is gapful excitation, and it 
touches the $\varepsilon_-$ branch at energy $3K$ 
on the six corners of the Brillouin zone. 
The $J=0$ case is 
special, since the mean-field state is an exact eigenstate, 
and the gapless excitations have 
a quadratic dispersion, $\varepsilon_{-,\mathbf{k}} \propto k^2$.  
The $J$=$K$ case is also special and the $\varepsilon_+$ branch 
becomes gapless and degenerate with the $\varepsilon_-$ branch.  

These two branches describe bosonic elementary excitations 
in the nematic order, and in particular, the gapless 
branch is the Goldstone mode corresponding to the broken spin 
rotation symmetry \cite{goldstone}.  These elementary excitations 
contribute to magnetic fluctuations, and therefore, 
we may call them {\em magnons} also in this case.  
The Gaussian approximation is checked by calculating 
$\rho$=
$ \langle \alpha_{\mathbf{r}}^\dagger \alpha_{\mathbf{r}}
 +\beta_{\mathbf{r}}^\dagger \beta_{\mathbf{r}} \rangle$, 
the density of mean-field excited states.  
$\rho (J)$ increases from 0 at $J$=0 to 
about 0.15 at $J$=$K$, which was much smaller than 1, 
and the approximation is justified semi-quantitatively. 

%%%%%%%%%%%%%%%%%%%%%%%%%%%%
% dynamical spin  structure factor
%%%%%%%%%%%%%%%%%%%%%%%%%%%%
%-------------------------------
\begin{figure}[t]
\begin{center}
\includegraphics[width=6cm]{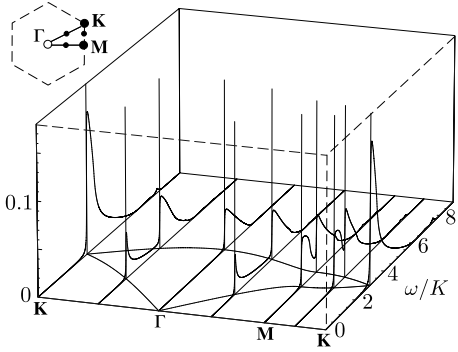}
\end{center}
\caption{Dynamical spin structure factor 
$\sum_{j} S_{jj}^{\mu \mu}(\mathbf{k},\omega)$
for $J/K$=0.5. Dispersion of two magnon 
branches is also plotted.}
\label{fig:sqomega}
\end{figure}
%-------------------------------
The dynamical spin structure factor is 
given at zero temperature by 
$S^{\mu \mu '}_{j j '}(\mathbf{k},\omega)$=
$\sum_\nu 
\langle 0 | S^{\mu}_{j,-\mathbf{k}} | \nu \rangle$ 
$\langle \nu | S^{\mu'}_{j',\mathbf{k}} | 0 \rangle $
$\delta(\mbox{$\omega$$-$$E_\nu$+$E_0$})$,  
where $|\nu \rangle$ is the eigenstate with energy $E_\nu$ and 
$|0 \rangle$ is the ground state.  
$S^{\mu}_{j,\mathbf{k}}$ is the Fourier transform 
of the spin on the $j$-sublattice. 
Structure factor $S^{\mu \mu '}_{j j '}(\mathbf{k},\omega)$ is 
diagonal with respect to spin indices $\mu$ and $\mu '$.
$\{ S^{xx}_{j j '} (\mathbf{k},\omega)\}$ is obtained as 
\begin{equation}
\left[ \begin{array}{ccc}
S^{xx}_{AA} & S^{xx}_{AB} & S^{xx}_{AC} \\
S^{xx}_{BA} & S^{xx}_{BB} & S^{xx}_{BC} \\
S^{xx}_{CA} & S^{xx}_{CB} & S^{xx}_{CC} 
\end{array}
 \right]
  = 
 \left[ \begin{array}{ccc}
 F_2 & 0 & 0 \\
 0 & F_1^{(0)} & e^{-i \phi_\mathbf{k}} F_1^{(1)} \\ 
 0 & e^{i \phi_\mathbf{k}} F_1^{(1)} & F_1^{(0)}
 \end{array}
 \right] , 
 \label{eq:matrixrep}
\end{equation}
where $e^{i \phi_\mathbf{k}}=\gamma_{\mathbf{k}}/|\gamma_{\mathbf{k}}|$.
$F_1^{(p)}(\mathbf{k},\omega)$ and $F_2(\mathbf{k},\omega)$ 
denote one and two-magnon contributions:
\begin{eqnarray}
F_1^{(p)}(\mathbf{k},\omega)
& =&  \frac{1}{2}\sum_{m=\pm}(-m)^p e^{2\theta_{m,\mathbf{k}}}
\delta(\omega-\varepsilon_{m,\mathbf{k}}),\\ 
F_2(\mathbf{k},\omega) &=& \frac{1}{4\Omega}
\sum_{\mathbf{q},m,n=\pm} \!
\sinh^2(\theta_{m,\mathbf{k+q}}-\theta_{n,\mathbf{q}}) \nonumber \\
& & \times 
\delta(\omega-\varepsilon_{m,\mathbf{k+q}}-\varepsilon_{n,\mathbf{q}}),
\end{eqnarray}
where  
$e^{4 m \theta_{m,\mathbf{k}}}$=
$(1+m \kappa |\gamma_{\mathbf{k}}|)/(1+ m|\gamma_{\mathbf{k}}|)$.
The summation of the momentum $\mathbf{q}$ is 
taken over the reduced Brillouin zone 
in the two-magnon contribution. 
$\{ S^{yy}_{jj'} (\mathbf{k},\omega)\}$ and 
$\{ S^{zz}_{jj'} (\mathbf{k},\omega)\}$ are 
given by
replacing sublattice indices $(A,B,C)$ in Eq.(\ref{eq:matrixrep})
by $(B,C,A)$ and $(C,A,B)$, respectively.
Figure \ref{fig:sqomega} shows 
$\sum_{j} S_{jj}^{\mu \mu}(\mathbf{k},\omega)$ for $J/K=0.5$. 
For each $\mathbf{k}$, there are two delta-function peaks 
at the one-magnon energies $\varepsilon_{\pm , \mathbf{k}}$, 
and they are accompanied by two-magnon continuum on the 
higher energy side.  
As $\mathbf{k} \rightarrow \mathbf{0}$, 
the delta-function peak of the gapless branch vanishes as 
$\sqrt{K/(8J)} |\mathbf{k}| \delta(\omega - \varepsilon_{-,\mathbf{k}})$. 
There are no magnetic Bragg peaks, consistent with the
absence of ordinary magnetic dipole order.

\begin{figure}[t]
\begin{center}
\includegraphics[width=6cm]{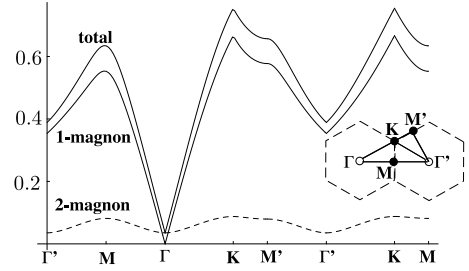}
\end{center}
\caption{Static structure factor 
$S(\mathbf{k})$ for $J/K$=0.5.} 
\label{fig:static}
\end{figure}

In Fig.~\ref{fig:static}, we present the static structure factor 
$S(\mathbf{k}) = 
\sum_{\mathbf{r}} \langle 0 | S^\mu(\mathbf{r})S^\mu(\mathbf{0})| 0 \rangle 
e^{i \mathbf{k}\cdot \mathbf{r}}$ for $J/K=0.5$. 
In the the original 
Brillouin zone of the triangular Bravais lattice, we have 
\begin{eqnarray}
S(\mathbf{k})&=&
{\textstyle \frac13}
\sum_{m=\pm} (1-m \cos \phi_{\mathbf{k}}) 
e^{2 \theta_{m,\mathbf{k}}} \nonumber \\
& &+{\textstyle \frac{1}{12\Omega}}
\sum_{\mathbf{q},m,n=\pm} \!
\sinh^2(\theta_{m,\mathbf{k+q}}-\theta_{n,\mathbf{q}}).
\end{eqnarray}
Note that we have $\phi_{\mathbf{k + \mathbf{b}_\pm}} = \phi_{\mathbf{k}} + 2\pi/3$, 
where  $\mathbf{b}_\pm=(\frac{2\pi}{3},\pm \frac{2\pi}{\sqrt{3}})$.
The one-magnon contribution of the gapful mode is strongly suppressed near $\mathbf{k}=\mathbf{0}$,
since we have $\cos \phi_{\mathbf{k}} \sim 
1 - k_x^2 (k_x^2-3 k_y^2)^2/1152$.
Around the $K$-point $\mathbf{k}_0=\frac13 (2\pi,2\pi/\sqrt{3})$, we have 
$e^{2\theta_{\pm,\mathbf{k}}} \sim 1-J/(2K)|\mathbf{k}-\mathbf{k}_0|$, 
and the $S(\mathbf{k})$ shows a singularity of cone tip.  
This leads to a power-law decay in the real-space 
spin-spin correlation as $|\mathbf{r}|^{-3}$. 
%-------------------------------
%-------------------------------

%%%%%%%%%%%%%%%%%
% magnetic susceptibility %
%%%%%%%%%%%%%%%%%
%-------------------------------
\begin{figure}[t]
\begin{center}
\includegraphics[width=6cm]{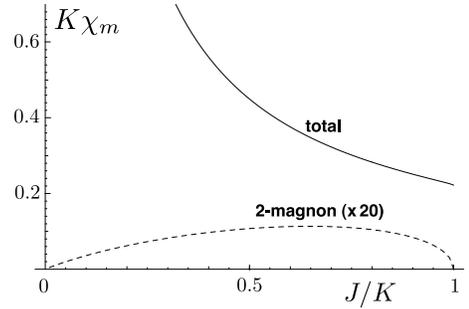}
\end{center}
\caption{Magnetic susceptibility $\chi_m$. 
Two-magnon contribution is multiplied 
by factor 20.}  
\label{fig:susceptibility }
\end{figure}
%-------------------------------
The magnetic susceptibility $\chi_m$ is calculated from the 
dynamical correlation function using the relation,
$\chi_m$=$\lim_{\mathbf{k} \rightarrow \mathbf{0}} \frac23 
\sum_{j,j ' \in \{ A,B,C \}} \int_0^\infty d \omega 
S^{\mu\mu}_{j j '}(\mathbf{k},\omega)/\omega$. 
As $\mathbf{k}$$ \rightarrow $$\mathbf{0}$, 
the one-magnon contribution of the gapful mode vanishes 
but that of the gapless mode converges to a finite value, 
$2/(9J)$.  Adding the two-magnon contribution, 
the result is 
\begin{equation}
\chi_m = \frac{2}{9J}+\frac{1}{3\Omega}\sum_{\mathbf{q}} 
\frac{\sinh^2(\theta_{+,\mathbf{q}}-\theta_{-,\mathbf{q}})}
{\varepsilon_{+,\mathbf{q}}+\varepsilon_{-,\mathbf{q}}},
\end{equation}
in physical units $(g\mu_B)^2$. It is noted that 
$\chi_m$ is isotropic in spin space.  
The one-magnon part is independent of $K$, and 
agrees with the classical value calculated 
by the mean-field approximation.  This value 
also coincides with the classical value for the 120-degree  
order in the pure Heisenberg model.  
In Fig.~\ref{fig:susceptibility }, $J$-dependence of $\chi_m$ 
is shown.  It is dominated by the one-magnon part and 
the two-magnon part is very small, 
about 1.85\% at largest.  
In both $J$=0 and $J$=$K$ cases, the two-magnon part of $\chi_m$ vanishes, 
and $\propto $$J$ around $J=$0 and 
$\propto $$\sqrt{K-J}$ around $J=K$.  

Similarly we can calculate the nematic correlation
$\langle {\mathcal Q}_{\mu \nu}(\mathbf{r}){\mathcal Q}_{\mu ' \nu '}(\mathbf{0}) \rangle$.
This tensor can be decomposed into
$
(1-\delta_{\mu \nu})(1- \delta_{\mu' \nu'})(\delta_{\mu \mu'}\delta_{\nu \nu'}+\delta_{\mu \nu'}\delta_{\nu \mu'}) 
G^{(1)}_{\mu \nu}(\mathbf{r})+
\delta_{\mu \nu} \delta_{\mu' \nu'} G^{(2)}_{\mu \mu'}(\mathbf{r})$.
The Fourier transform $G^{(1)}_{\mu \nu}(\mathbf{k})$ has a similar structure  as in  Eq.(\ref{eq:matrixrep}),   
which consists of one and  two-magnon parts, while  $G^{(2)}_{\mu \mu'}(\mathbf{k})$
has the delta-function peaks
reflecting the existence of static nematic
quadrupole moments and two-magnon part.
The one-magnon part of the gapless mode in $G^{(1)}_{\mu \nu}(\mathbf{k})$
diverges as $|\mathbf{k}|^{-1}$ in the limit $\mathbf{k} \rightarrow \mathbf{0}$. 
The two-magnon part diverges more slowly as $\log|\mathbf{k}|$ around $\mathbf{k}= \mathbf{0}$
in both  $G^{(1)}_{\mu \nu}(\mathbf{k})$ and $G^{(2)}_{\mu \mu'}(\mathbf{k})$. 

Let us discuss the implications of the above results 
and compare with the experiments for NiGa$_2$S$_4$ \cite{nigas}.  
We have found gapless bosonic excitations with linear 
energy dispersion.  They contribute to specific heat as, 
$C$$\sim12\pi \zeta(3) (T/v)^2$$\sim45.3 (T/v)^2$ 
in units of $k_B$ per spin,  
where $v$=$\sqrt{9JK/2}$ is the velocity of the gapless mode.  
Since the order parameter is a tensor with continuos 
degrees of freedom as for the 120-degree magnetic order \cite{kawamura}, 
the nematic order does not appear at finite temperatures 
in two-dimensional systems \cite{mermin}.  
The zero-temperature magnetic susceptibility is finite 
and given by $\chi_m \sim 2/(9J)$.  
These behaviors agree with the experimental data.  
The spin structure factor does not show 
magnetic Bragg peaks, implying the absence of 
ordinary magnetic long-range order.  
As for the static spin structure factor $S(\mathbf{k})$, 
the three-sublattice nematic order shows a sharp peak 
at the corners of the reduced Brillouin 
zone that are inside the original 
Brillouin zone of the triangular Bravais lattice, 
and the spin-spin correlation function shows a power-law decay 
in space $\langle S(\mathbf{r}) S(\mathbf{0}) \rangle 
\sim 1/|\mathbf{r}|^3$.  
We should emphasize that the peak in $S(\mathbf{k})$ 
does not diverge but has only a kink singularity, 
$S(\mathbf{k}) \sim \mbox{const}-A |\mathbf{k} - \mathbf{k}_0|$.  
In neutron experiments, the spin correlation length was
determined as the inverse of the peak width, and 
therefore that is consistent with our result.  
There remain two points that should be further investigated 
in future studies.  The first point is a detailed structure of 
$S(\mathbf{k})$.  The peak position is different between the neutron data
and the theoretical calculation shown before.
We believe that it is possible to reproduce similar
$S(\mathbf{k})$ by tailoring the starting Hamiltonian by including
longer-range interactions, like in the case of incommensurate
helical spin order.  We should emphasize that
the basic features presented above will not change
aside from detailed structure in $S(\mathbf{k})$.  The second point is 
the energy dependence. 
We also need more quantitative analysis on 
the physical properties at finite temperatures 
and also under finite magnetic fields, but 
we believe that the scenario presented in this paper 
will help the understanding of the intriguing properties 
of NiGa$_2$S$_4$.  

The authors thank Satoru Nakatsuji, Yusuke Nambu, 
and Tsutomu Momoi for stimulating discussions.  
This work was supported under a Grant-in-Aid for 
Scientific Research from the Ministry of Education, 
Science, Sports, and Culture of Japan.


\begin{thebibliography}{99}
\bibitem{momoi}
One can also define similar nematic order parameters  
for spin-1/2 systems, if we compose effective spin 
$S \ge 1$ by multiple spins with $S=1/2$.
\end{thebibliography}
\end{document}